\documentclass[pra,aps,twocolumn,showpacs,floatfix,superscriptaddress]{revtex4}
\usepackage{amssymb,amsfonts
}
\usepackage{times}
\usepackage{graphicx}
\usepackage{bm}
\usepackage{amsmath}
\usepackage{ulem}
\usepackage{color}

\makeatletter
\def\NAT@bibsetnum#1{%
 \setlength{\topsep}{\z@}%
 \NATx@bibsetnum{#1}%
}%
\renewenvironment{thebibliography}[1]{%
 \NAT@thebibliography{#1}%
 \@clubpenalty\clubpenalty
 \let\@TBN@opr\present@bibnote
 \@FMN@list
}{%
 \@endnotesinbib
 \edef\@currentlabel{\arabic{NAT@ctr}}%
 \NAT@endthebibliography
 \global\let\auto@bib\@empty
}
\newcommand*{\supplementarystart}{%
  \close@column@grid%
  \clearpage%
  \onecolumngrid%
  \setcounter{enumiv}{0} 
  \setcounter{equation}{0} 
  \setcounter{figure}{0} 
  \setcounter{table}{0} 
  \setcounter{page}{1}
  \c@secnumdepth=4
  \renewcommand{\theequation}{s\arabic{equation}} 
  \renewcommand{\bibnumfmt}[1]{[s##1]} 
  \renewcommand{\@onlinecite}{s\citealp} 
  \renewcommand{\cite}[1]{{[}\onlinecite{##1}{]}}
  \renewcommand{\thefigure}{s\arabic{figure}}
  \renewcommand{\thetable}{s\Roman{table}}
  \renewcommand{\thepage}{s\arabic{page}}
}
\makeatother

\newcommand{\be}{\begin{equation}}
\newcommand{\ee}{\end{equation}}
\newcommand{\bea}{\begin{align}}
\newcommand{\eea}{\end{align}}
\newcommand{\p}{\partial}
\newcommand{\BEA}{\begin{eqnarray}}
\newcommand{\EEA}{\end{eqnarray}}
\newcommand{\BC}{\begin{cases}}
\newcommand{\EC}{\end{cases}}
\newcommand{\beml}{\begin{subequations}}
\newcommand{\eml}{\end{subequations}}

\newcommand{\bmp}{\begin{minipage}}
\newcommand{\emp}{\end{minipage}}
\newcommand{\epm}{\end{pmatrix}}
\newcommand{\bpm}{\begin{pmatrix}}
\newcommand{\ba}{\begin{array}}
\newcommand{\ea}{\end{array}}
\newcommand{\lt}{\left}
\newcommand{\rt}{\right}

\DeclareMathOperator{\im}{Im}

\begin{document}
\title{Plasmonic shock waves and solitons in a nanoring}

\author{K.\,L.~Koshelev}
\affiliation{A.\,F.~Ioffe Physico-Technical Institute, 194021 St.~Petersburg, Russia}
\affiliation{ITMO University, 197101 St.~Petersburg, Russia}
\affiliation{L.\,D.~Landau Institute for Theoretical Physics, Kosygina street 2, 119334 Moscow, Russia}

\author{V.\,Yu.~Kachorovskii}
\affiliation{A.\,F.~Ioffe Physico-Technical Institute, 194021 St.~Petersburg, Russia}
\affiliation{L.\,D.~Landau Institute for Theoretical Physics, Kosygina street 2, 119334 Moscow, Russia}
\affiliation{ Rensselaer Polytechnic Institute,  110, 8$^{th}$ Street, Troy, NY, 12180, USA}

\author{M.~Titov}
\affiliation{Radboud  University,  Institute  for  Molecules  and  Materials,  NL-6525  AJ  Nijmegen, The  Netherlands}

\author{M.\,S.~Shur}
\affiliation{Center for Integrated Electronics, Rensselaer Polytechnic Institute, 110, 8$^{th}$ Street, Troy, NY, 12180, USA}

\begin{abstract}
We apply the hydrodynamic theory of electron liquid  to demonstrate  that a circularly polarized radiation induces the diamagnetic, helicity-sensitive \textit{dc} current in a ballistic nanoring. This current is dramatically enhanced in the vicinity of plasmonic resonances. The resulting magnetic moment of the nanoring represents a giant increase of the inverse Faraday effect. With increasing radiation intensity, linear plasmonic excitations  evolve into  the strongly non-linear plasma shock waves. These excitations  produce a series of the  well resolved  peaks  at   the THz frequencies.   We demonstrate that the plasmonic wave dispersion transforms the shock waves into solitons. The predicted effects should enable multiple applications in a wide frequency range (from the microwave to terahertz band) using optically controlled ultra low loss electric, photonic and magnetic devices.
\end{abstract}


\maketitle
\pagebreak[100]

\section{Introduction}
The feature size of modern electronic and photonic devices has dropped down to 10\,nm. At such scales plasmonic excitations become a salient feature determining the device performance. This explains a recent surge of interest to plasmonics \cite{Rev1,Rev2,Rev3,Rev4,Rev5,Rev6,Rev7}, the field which has to be further explored from the fundamental physics point of view \cite{Rev7}.

Much of plasmonic physics can be captured by the hydrodynamic approach that is becoming increasingly relevant for electronic and spintronic devices due to fast improving quality of nanostructures. The first theories and measurements of the hydrodynamic effects on charge transport date to the early work by Gurzhi \cite{Gurzhi} and by Jong and Molenkamp \cite{Molenkamp}. In recent years the field received a revived attention driven by the development of high-mobility nanostructures \cite{Jaggi, h1,h2,h3,h4,h5,h6,h7,h8} and graphene \cite{h9,h10,h11,h12,h13,h14} where the electron-electron collision-dominated transport regime can be reached.

The interest to non-linear plasmon waves has been stimulated in early 90s by exploring the analogy between the ``shallow water'' hydrodynamics and that of the electron
liquid in two-dimensional (2D) gated systems. It was shown that the electron liquid in these systems could become unstable with respect to the excitations of tunable plasma oscillations \cite{1}.  Many other beautiful hydrodynamic phenomena such as choking of electron flow \cite{2}, nonlinear rectification of plasma waves \cite{3,4} and formation of plasmonic shock waves \cite{5} have been subsequently proposed. Possible applications of these phenomena to plasma-wave electronics were intensely discussed (see the reviews~\cite{Rev8,Knap}). In particular, much attention has been paid recently to the generation of plasmonic oscillations in the field-effect transistors (FETs) for realizing tunable THz emitters or detectors \cite{Rev8,Knap}.  

The detector responsivity is enhanced dramatically in the presence of \textit{dc} current \cite{Veksler}.  It can be also enhanced by making artificial periodic structures such as FET arrays and periodically grated gates \cite{Aizin1, Azin2, my1, Azin3, Wang1}. Such plasmonic crystals have already demonstrated excellent performance as THz detectors  \cite{allen1,allen2,allen4,28,allen6} in a good agreement with the theory \cite{32,30,31,popov}.  Moreover, THz emission from grating gate structures have been also recently  reported \cite{29,new3}.

Having a non-zero \textit{dc} photovoltaic response in a single FET requires an inversion asymmetry which may be created by boundary conditions \cite{1}.  Plasmonic crystals  would require an inversion asymmetry within the unit cell of a crystal. Such an asymmetry can be induced by a ratchet effect (see the review paper \cite{Ivchenko2011} and the references therein). The latter is also strongly enhanced by plasmons \cite{ratchet-my}.

Here we explore another system enabling a greatly enhanced coupling between THz radiation and plasmonic excitations -- a ballistic nanoring.  Such a system has a number of advantages compared to a single FET.  First of all, an inversion asymmetry is not required in this case  because of the nanoring multi-connected geometry \cite{Kibis11,Kibis13,Alexeev13,Joibari14,Alexeev12,Kruglyak2005,Kruglyak2007,Koshelev2015} .

More importantly, we now predict that the plasmonic resonances in a high-quality nanoring can be much sharper as compared to a FET. Indeed, the dissipation in contacts and the coupling to ungated regions in the FET leads to essential weakening of plasmonic resonances. In a nanoring these deleterious effects may be fully avoided while the coupling can be further enhanced by fabricating arrays of identical nanorings.

\begin{figure}[t]
\center{\includegraphics[width=\columnwidth]{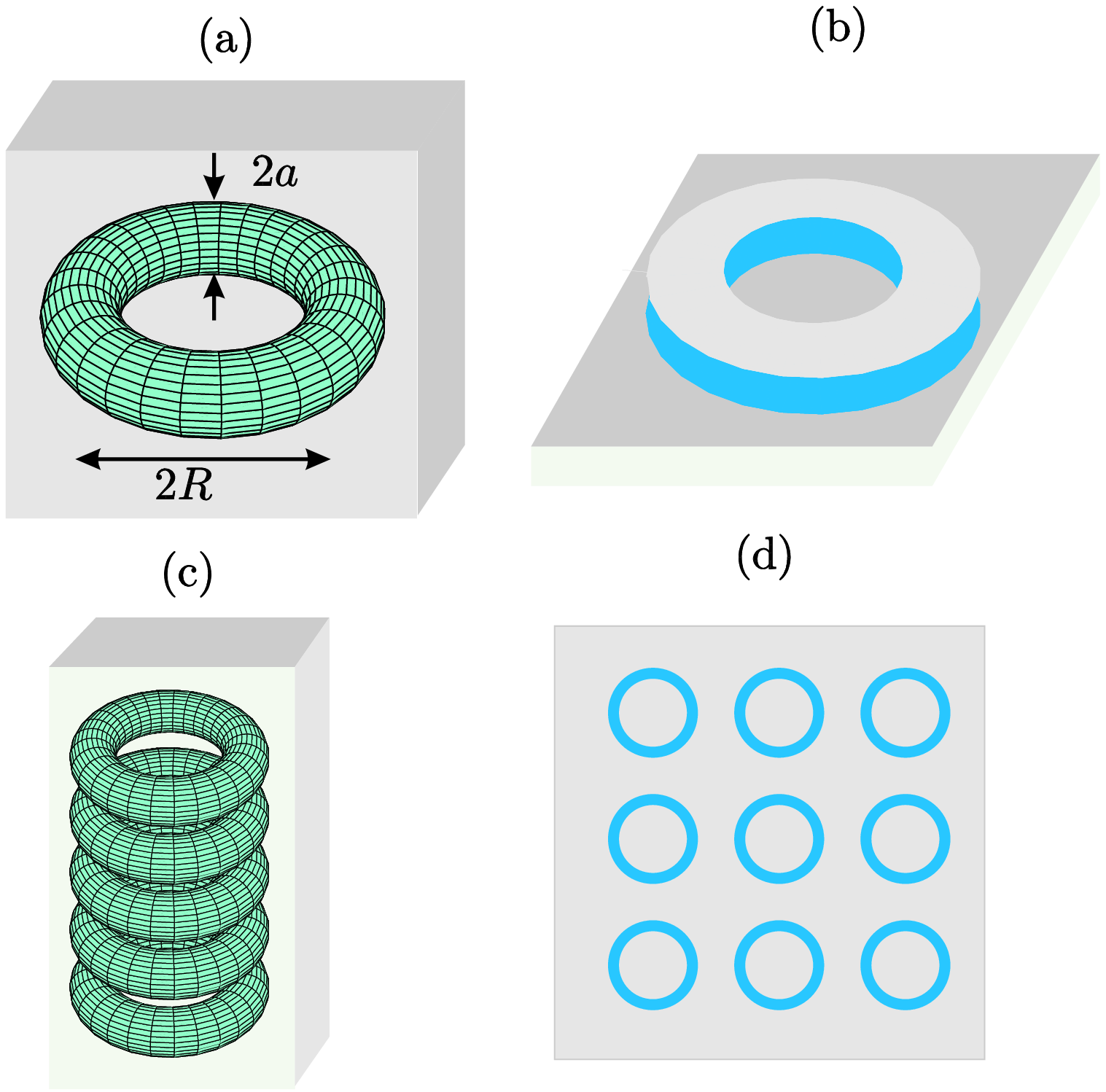}}
\caption{Three dimensional (a) and two-dimensional (b) quantum nanorings. Three-dimensional ring-stacked arrays (c) and two-dimensional arrays of rings (d) on a substrate.}
\label{Fig1}
\end{figure}

In Fig.\,\ref{Fig1} we illustrate possible realizations of nanorings and nanoring arrays.  A  quasi-onedimensional (1D) ring  can be  fabricated from 2D or 3D metals or semiconductors as shown in Figs.\,\ref{Fig1}a-\ref{Fig1}b, respectively. The arrays of nanorings made of these materials are depicted schematically in Figs.\,\ref{Fig1}c-\ref{Fig1}d .  

Plasmonic excitations in both 2D and 3D types of nanorings are nearly identical due to similar electrostatic properties of these quasi-1D systems. Still, it is much easier to produce clean rings made of 2D semiconductor materials. Such rings can be fabricated with the use of standard semiconductor technology: by growing a narrow 2D semiconductor on a substrate followed by patterning a nanoring or an array of nanorings. One can also use a gate electrode (or an array of gate electrodes) to control electron concentration in the nanoring. 

Below we demonstrate that sharp plasmonic resonances can be excited in semiconductor GaAs or GaN nanorings within a wide range of sizes and carrier concentrations assuming realistic values of electron mobility and a reasonable temperature range. Similar effects can be observed in rings made of graphene and in systems of different geometry such as self-assembled nanorods or nanodisks. The difference in the latter case would only concern somewhat more complex electrostatics of such systems.

We predict that excitation of plasmonic waves by circularly polarized radiation leads to a resonant optical rectification effect -- a large diamagnetic circulating \textit{dc} current that manifests itself as a magnetic moment of the nanoring. When radiation intensity exceeds a critical value, the plasmonic waves transform into shock waves (SW) that might further develop into multiple solitons (a similar effect was recently predicted for nonlinear waves in the Luttinger liquid \cite{prot1,prot2}). In this regime, the system is functioning as an efficient emitter of high frequency radiation harmonics. One possible application of the plasmonic SWs is to transform circularly polarized resonant GHz waves into a number of well resolved peaks at THz frequencies. 

Circulating current in a nanoring gives rise to the inverse Faraday effect (IFE), which is the excitation of helicity-sensitive magnetic moment by a circularly polarized light \cite{Pitaevskii61,Ziel65,Kimel05,belotelov}. The IFE has been widely discussed in connection with ultrafast magnetization dynamics \cite{Kimel05,belotelov,Kirilyuk10,Kirilyuk11}. The phenomenon is closely related to the quantum IFE in nanorings \cite{Kibis11,Kibis13,Alexeev13,Joibari14,Alexeev12,Kruglyak2005,Kruglyak2007,Koshelev2015} and in a chaotic cavity \cite{Polianski2009}  as well as to the optical analog of Aharonov-Bohm effect for excitons in a semiconductor quantum ring \cite{kibis15}.  Remarkably, the plasmonic IFE described below is based on quasiclassical mechanism  and, consequently, orders of magnitude stronger than the corresponding quantum phenomenon.

We would like to stress that the closed rings, we consider, have an important advantage as compared to the ring-split resonators (see Ref.~\cite{ring-split} and reference therein).  The latter can create large values of optically induced {\it alternating} magnetic field but can not conduct circulating \textit{dc} current, and, consequently, do not produce a {\it constant} magnetic field. In contrast, excitation of a closed ring by a circularly polarized radiation may produce a sizable constant magnetic field (of the order of a Gauss for a single nanoring). Similar phenomena should occur in metallic films perforated with hole arrays \cite{belotelov}.

\section{Model}
In this paper, we discuss excitations of plasmonic resonances in a single nanoring. Generalization for the case of ring array is straightforward. We consider two basic setups: i) a nanoring of the radius $R$ made of 3D wire with a diameter $2a$  (see Fig.\,\ref{Fig1}a) and ii) a nanoring of the same radius made of 2D strip with the width $2a$ (see Fig.\,\ref{Fig1}b). We assume that the nanorings are subject to a circularly polarized electromagnetic radiation with electric field parallel to the ring plane. The radiation wavelength is assumed to be much larger than $R$, so that electric field is homogeneous within the ring size. At the same time we naturally assume $R\gg a \gg \lambda_{\rm F},$ where $\lambda_{\rm F}$ is the electron Fermi wavelength. In this case, the ring is multichannel and can be described quasiclassically, while at the same time it can be considered as a quasi 1D wire from the electrostatics point of view. Under these assumptions, the electric field induced by plasma wave can be expressed in terms of linear electron concentration (concentration per unit length) for both types of rings. 

Plasmonic resonances take place in high-quality multi-channel nanorings where electron-electron collisions dominate over scattering off phonons and impurities. The latter condition may be formulated as $\tau_{\rm ee} \ll \tau_{\rm tr}$, where $\tau_{\rm ee}$ is the electron-electron collision time, while $\tau_{\rm tr}$ stands for the transport scattering time. The condition ensures the validity of the hydrodynamic approach. 

The hydrodynamic equations, describing electron liquid in a multi-channel nanoring, can be derived in a standard way from kinetic equation assuming that the electron distribution function depends only on hydrodynamic parameters, i.\,e. on the local electron density, local velocity and local temperature. The derivation (for the case of 2D systems) can be found, e.\,g., in Ref.~\cite{ratchet-my}. Neglecting heating effects (see. Ref.~\cite{ratchet-my}) and integrating the hydrodynamic equations derived in Ref.~\cite{ratchet-my} over the ring cross-section, one arrives at the hydrodynamic equations for the linear electron concentration $N$ and the hydrodynamic velocity $V$,
\begin{align}
\label{N}
&\frac{\p N}{\p t}+\frac{\p\left(N V\right)}{\p x }=0,\\
\label{V}
&\frac{\p V}{\p t}+ V\frac{\p V}{\p x } - \eta \frac{\p^2V}{ \p x^2} = -\gamma V  -\frac{\p \Phi}{ \p x}+\frac{eE_0 }{m\varepsilon} \sin\theta,
\end{align}
where $x$ is the coordinate along the ring, $E_0$ is the amplitude and $\omega$ is the frequency of circularly polarized radiation, $\eta$ is the kinematic viscosity of electron liquid, $m$ is the effective electron mass,  $\gamma=1/\tau_{\rm tr}$ is the friction due to scattering off impurities and phonons, $\varepsilon$ is the dielectric constant, and the angle $\theta$ is defined as 
\be\theta=x/R-\omega t.\ee  For a ring made of 3D material, the derivation is fully analogous and yields the same system of equations.

The electrostatics of a thin nanoring is solved by the following potential (see  Appendix~\ref{appA})
\be
\label{Phi}
\Phi\!=\! \frac{e^2}{m\varepsilon}\!\left [\! (N\!-\!N_0) \Lambda +d^2 \frac{\p^2N}{\p x^2}\right]\!=\! s^2
\left[n\!+\! \frac{d^2}{\Lambda} \frac{\p^2 n}{\p x^2}\right],
\ee
where $\Lambda =\ln(d^2/a^2)$,  $N_0$ is the linear (1D) charge concentration in equilibrium, $d$ is the screening radius ($a\ll d \ll R$), 
\be n=(N-N_0)/N_0 \ee is the relative dimensionless concentration, and
\be s=\sqrt{\frac{e^2N_0 \Lambda}{m\varepsilon }} \label{s} \ee
is the plasma wave velocity, which might be tunable by the gate voltage.

Two possible experimental realizations of the nanoring discussed above give rise to $N_0 = \pi a^2 n_{\rm 3D}$ for 3D wire and $N_0 = 2a n_{\rm 2D}$ for 2D wire, where
$n_{\rm 3D}$  ($n_{\rm 2D}$) is the equilibrium value of 3D (2D) electron concentration. One should also specify the dielectric constant entering Eqs.~\eqref{Phi} and \eqref{s}.   For rings made of 3D wires, $\varepsilon$ is given by the dielectric constant of the material in which the ring is embedded.  For a ring made from 2D strip, sandwiched between two materials having dielectric constants $\varepsilon_1$ and $\varepsilon_2$, the effective dielectric constant is given by $\varepsilon = (\varepsilon_1+\varepsilon_2)/2$.  For example, for a 2D ungated ring placed on the surface between vacuum (or air) and substrate with the dielectric constant $\varepsilon_1$ one gets  $\varepsilon =(\varepsilon_1 + 1)/2$.

In Eq.~\eqref{Phi}, we neglect the pressure of electron liquid assuming that $s$ is large as compared to the Fermi velocity. We also neglect all thermoelectric forces (as compared to $\p\Phi/\p x$) thus decoupling Eqs.~\eqref{N} and \eqref{V} from the heat equation \cite{ratchet-my}. Finally, we neglect  the dependence of $\eta$ on $N$ (setting $\eta(N) \approx \eta[N_0]$) and regard $N$ to be smooth on the scale of $d$, thus keeping the main logarithmic contribution to the Coulomb potential and the leading correction to it (see Appendix~\ref{appA}). The latter describes a weak plasmonic dispersion. The remaining subtlety concerns boundary conditions at the surface of the ring.  Frequently used no-slip condition, $V=0$, would result in the Poiseuille flow and, consequently, in a relatively large resistance caused by viscosity. On the other hand, recent technology allows for fabricating quantum wires and rings of an extremely high quality. This implies that the friction originating at the surface of the ring might be certainly too low to drive the ring into the Poiseuille regime. In our derivation of Eqs.~\eqref{N}, we fully neglect this boundary-induced friction thus making $N$ and $V$ depend only on $x$.  A more general case of arbitrary strong surface friction is briefly discussed in Appendix~\ref{appB}.

\section{Linear  regime}
When the radiation intensity is small, the Eqs.~(\ref{N},\ref{V}) can be linearized. In the absence of radiation and for $\eta=\gamma=0$, plasma waves propagating in a ring have simple linear spectrum
\be
\omega(k) =s k,
\label{spektr}
\ee
where $s$ is given by Eq.~\eqref{s} [here we neglect the small dispersion due to the second term in the square brackets in Eq.~\eqref{Phi} ]. The wave vectors are quantized: \be k_n=n/R,\ee where $n$ is the integer number ($n \neq 0$).  Finite friction $\gamma$ and viscosity $\eta$ would lead to damping of plasma waves that is similar to damping effects in FETs \cite{1}.

A weak external radiation field impinging on the ring couples to the electronic fluid and excites linear plasmonic oscillations with the fundamental frequency
\be
\omega_0=\frac{s}{R}=\frac{a}{R}\sqrt{\frac{\pi e^2 \Lambda n_{\rm 3D}}{m\varepsilon}}.
\label{omega0}
\ee
For a circularly polarized radiation, the oscillations are rectified to produce the \textit{dc} circulating current that peaks at the plasmonic resonant frequencies: 	
\be
I_{\rm dc} = e\langle NV\rangle,
\label{averaged-I}
\ee
where the brackets stand for the time average. The direction of the current is determined by the radiation helicity (below we put $\omega >0$):
\be I_{\rm dc} (\omega)= - I_{\rm dc} (-\omega).\ee
We now introduce the rescaled quantities
\BEA
&& J=I_{\rm dc}/ eN_0 R, \qquad v=V/R,
\label{rescaled1}
\\
&& \varkappa=\eta/R^2, \qquad \beta=\omega_0^2 d^2/\Lambda\omega R^2,
\label{rescaled2}
\EEA
that we respectively refer to as current, velocity, viscosity and dispersion.

Solving the linearized equations, we find in the resonance approximation, i.\,e.\ for  $\delta=\omega_0-\omega \ll \omega_0$ that
\BEA
&&v =\omega n= \frac{eE_0}{mR\varepsilon} \im \frac{e^{i \theta}}{\varkappa+\gamma +i ( 2\delta -\beta)},
\label{vv}
\\
&&J \!=\!\langle nv\rangle \!=\!\frac{1}{2\omega} \left(\frac{eE_0}{ m R\varepsilon}\right)^2 \!\frac{1}{(\varkappa+\gamma)^2+ (2\delta -\beta)^2},
\label{J}
\EEA
where $\beta \approx \omega_0 d^2/\Lambda R^2$  for $\delta\ll\omega_0$. Thus, the {\it dc} response has a Lorentzian shape that peaks at the plasmonic frequency with a small dispersion-induced shift $\beta$ and is broadened by disorder and viscosity.


The key condition for observation of sharp plasmonic resonance is a sufficiently high quality factor. This factor is determined by viscosity, dispersion, and disorder (and/or phonon) scattering. The resonances are sharp provided that $\omega_0 /\varkappa  \gg 1$, $\omega_0 /\beta  \gg 1$,  and  $\omega_0 \tau_{\rm tr} \gg 1$. Since the plasma wave frequency $\omega_0$ decreases with the ring radius $R$, the conditions above yield the upper bound for $R$. The low bound for the ring radius (at fixed ratio $R/a$)   is determined by the Fermi-wave length since the ring has to support a large number of quantum channels. (In a single channel ring one should take into account Luttinger liquid effects, but the qualitative predictions of our theory will be still valid. A more formal analysis of the Luttinger liquid rings may be developed along the lines of Refs.~\cite{prot1,prot2}.) In Sec.~\ref{discussion}, we demonstrate that all three parameters $\omega_0 /\varkappa,\omega_0 /\beta, $ and $\omega_0 \tau_{\rm tr} $ might be simultaneously large (of the order of $10 \div 100$) for realistic semiconductor rings with a large number of quantum channels.

\begin{figure}[t]
\centerline{\includegraphics[width=0.98\columnwidth]{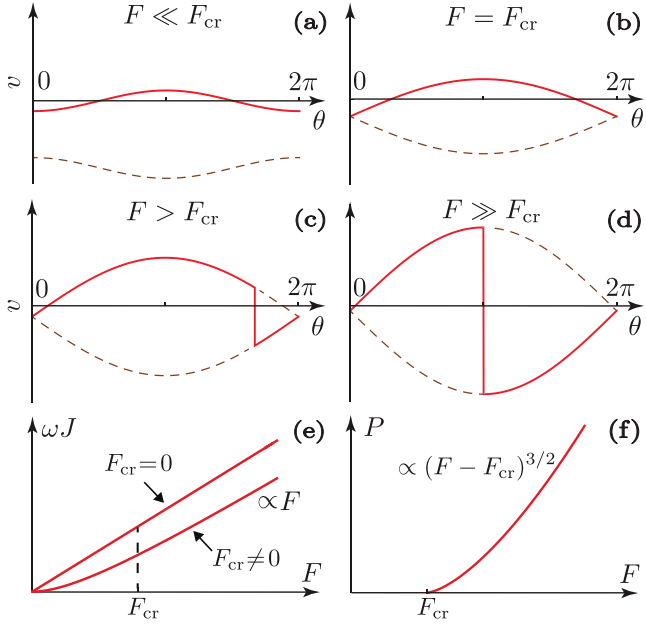}}
\caption{Numerical solution of Eqs.~(\ref{N},\ref{V}) showing velocity $v$ and concentration $n\approx v/\omega$ profiles for different values of $F$ (a-d). Solid and dashed lines in (a,b) correspond to positive and negative sign choice in Eq.~\eqref{q-tau}. At $F=F_{\rm cr}$ these solutions touch each other (b). Critical amplitude $F_{\rm cr}$ corresponds to a formation of the SW front. Panels (c) and (d) illustrate the numerical solution for $F>F_{\rm cr}$ that experiences a jump between positive and negative branches (the two solutions of Eq.~\eqref{q-tau} are indicated with the dashed lines). The panels (e) and (f) show the dependence of the current $J$ and the dissipated power $P$ on $F$ for $\beta=0$ and $\varkappa \to 0$.}
\vspace*{-0.3cm}
\label{Fig2}
\end{figure}
%
\begin{figure}[t]
\centerline{\includegraphics[width=0.98\columnwidth]{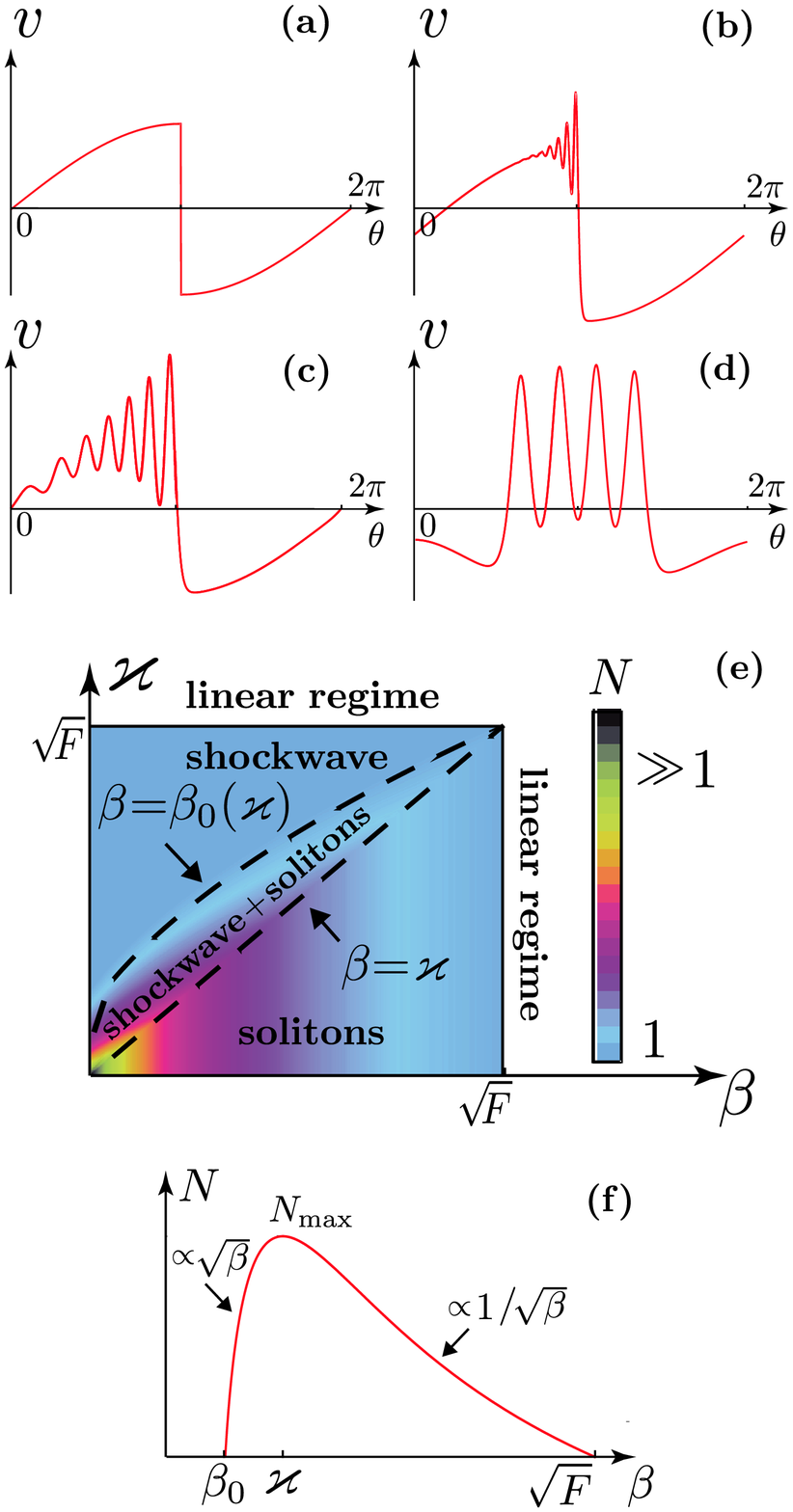}}
\caption{Evolution of the solution of Eqs.~(\ref{N},\ref{V}) that demonstrates the emergence of solitons at the SW front with increasing dispersion coefficient $\beta$ at $F\gg F_{\rm cr}$ and $\varkappa \ll \sqrt F$:  (a) $\beta \ll\beta_0$, (b) $\beta \gtrsim\beta_0$,  (c) $ \beta \sim \varkappa$,  (d) $ \varkappa/\delta\tau \gg \beta \gg \varkappa$. The number of solitons $N$ as a phase diagram in the dispersion-viscosity plane (e) and the dependence of $N$ on the dispersion parameter $\beta$ (f).}
\vspace*{-0.3cm}
\label{Fig3}
\end{figure}

\section{ Nonlinear regime}

For larger radiation intensities, the non-linear terms in Eqs.~(\ref{N},\ref{V}) become increasingly important. Figs.\,\ref{Fig2} and \ref{Fig3} show the results of the numerical analysis of Eqs.~(\ref{N},\ref{V}) using a finite element method (see Appendix~\ref{appE}). We find that, at sufficiently long times, the solution is stationary in the rotating reference frame. No chaotic or turbulent behavior is obtained. The results obtained numerically at long times can be reproduced analytically by analyzing the automodel solutions with $n=n(\theta)$, $v=v(\theta)$ that satisfy the neutrality condition $\langle  n \rangle=0$, where the angular brackets now stand for the averaging over the angle $\theta$. In this case, the Eq.~\eqref{V} imposes the constraint
\be \langle v\rangle =0.\ee
One may integrate the Eq.~\eqref{N} with the help of the constraint to obtain
\be J =-\omega n +(1+n) v.\ee
For sufficiently small velocities, $v\ll\omega$, one finds the charge density
\be n=({J-v})/({v-\omega})\approx  {v}/{\omega} +{v^2}/{\omega^2} -{J}/{\omega},\ee
which is substituted into Eq.~\eqref{N} to obtain a useful equation
\be
\frac{\p}{\p \theta}\! \left [2 v \delta \!+\!\frac{3}{2} v^2\! -\!\varkappa \frac{\p v}{\p \theta}  +\beta \frac{\p^2 v}{\p \theta^2}\!    \right]\!=\! \!-\!\gamma v\!+\!\frac{eE_0}{m R\varepsilon}\sin\theta.
\label{dv}
\ee
that holds in the resonant approximation. The electric current $J$ is found from the solution of Eq.~\eqref{dv} as 
\be J=\langle nv \rangle  \approx {\langle v^2\rangle }/{\omega}.\ee

Since both the viscosity and disorder suppress resonant behavior in a similar fashion [see Eq.~\eqref{J}], we consider, for simplicity, the case $\gamma=0$.  (Importantly, the limit $\gamma \to 0$ should be taken with care, since any small but finite $\gamma$ guarantees the constraint $\langle v\rangle=0$ that follows immediately from averaging Eq.~\eqref{dv} over the angle $\theta$. In what follows we neglect the term $\gamma v$ in Eq.~\eqref{dv} but respect the constraint.) We integrate Eq.~\eqref{dv} over the angle and introduce the variables
\be
q=3v/2 +\delta,\quad   F= 3eE_0/2mR\varepsilon
\ee
 to find
\be
\beta \ddot{q} +\varkappa \dot{q}=q_0^2-q^2-F\cos\tau, \quad (\tau=-\theta)
\label{q}
\ee
where \be q_0^2=\langle q^2\rangle\ee is the integration constant, which has to be found self-consistently, and $\dot{q}\equiv dq/d\tau$. The Eq.~\eqref{q} coincides with the Newton equation of motion for a particle with the "mass" $\beta$ oscillating in a classical cubic potential \be U(q)={q^3}/{3} -q_0^2 q \ee under the action of both the external dynamic force $-F\cos\tau$ and the "friction force" $-\varkappa \dot{q}$. The motion is further constrained by two conditions, 
\BEA
q(\tau)&=&q(\tau+2\pi),\\
\langle q\rangle&=&\delta.
\EEA
The potential $U(q)$ has two stationary points (see Fig.\,\ref{Fig4}):  $q=q_0$ (stable minimum) and  $q=-q_0$ (unstable maximum) with the corresponding energies given by $U(q_0)=-2q_0^3/3$ and $U(-q_0)=2q_0^3/3$. For small values of $F$, the particle undergoes linear oscillations around the stable point. Expanding $q_0^2-q^2\approx 2 q_0 (q_0-q)$ in the r.\,h.\,s. of Eq.~\eqref{q} and solving the corresponding linear equation one readily reproduces Eq.~\eqref{vv}. In this case, $q_0 \approx \delta$.

Let us fix $\delta$ at a certain value and increase $F$ to drive the system into a nonlinear regime. First, we assume that both viscosity and dispersion are absent ($\varkappa=\beta=0$). In this case the Eq.~\eqref{q} has two solutions
\be
q(\tau) =  \pm \tilde q_0 (\tau), \,\,\,\, \tilde{q}_0 (\tau) =\sqrt{q_0^2-F\cos\tau},
\label{q-tau}
\ee
where $\tilde{q}_0 (\tau)$ stands for a position of extremum of the dynamic potential \be \tilde U=U(q)+Fq\cos\tau= {q^3}/{3} -\tilde q_0^2 q.\ee 
Since $\langle q \rangle =\delta$, the choice of the right solution is fixed by the sign of $\delta$. To be specific we let $\delta >0$ below. Upon angle averaging the dependence of $q_0$ on $F$ and $\delta$ is given implicitly by
\be
\delta = \int\frac{d\tau}{2\pi} \sqrt{q_0^2-F\cos\tau}.
\label{aver}
\ee
This equation has a solution only for $F<F_{\rm cr}$, where
\be
F_{\rm cr}=\pi^2\delta^2/8.
\ee
The linear regime is reproduced in the limit $F\ll F_{\rm cr}$ (see Appendix~\ref{appC1}). The corresponding solution for $v(\theta)$ is shown by a solid line in Fig.\,\ref{Fig2}a. The dashed line corresponds to the choice of minus sign in Eq.~\eqref{q-tau}. For $F>F_{\rm cr}$ the result of Eq.~(\ref{aver}) breaks down and the velocity profile is discontinuous (detailed calculation is relegated to Appendix~\ref{appC1}), i.\,e. a step (SW front) appears at a certain point $\tau=\tau_0$.  The amplitude of the step is given by $2\tilde q_0 (\tau_0)$, where
\be
\cos(\tau_0/2)=\sqrt{F_{\rm cr}/F},\quad \tilde q_0 (\tau_0)= \sqrt{2(F-F_{\rm cr})}.
\label{cos}
\ee
Note that the amplitude of the SW front increases monotonously with $F$ and is given by $\sqrt{8F}$ in the limit $F \gg F_{\rm cr}$. In this limit, the front is located at $\tau_0\approx \pi$ (see Fig.\,\ref{Fig2}d).

\subsection{Finite viscosity}  

\begin{figure}[t]
\center{\includegraphics[width=0.9\columnwidth]{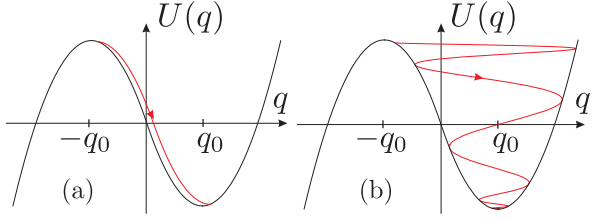}}
\caption{The effective potential (black lines) and solutions (red lines) of Eq.~\eqref{q} for $\beta \ll \beta_0$ and (a) for $ \beta_0<\beta <\varkappa$  (b) for $F>F_{\rm cr}$. }
\label{Fig4}
\end{figure}

Let us now switch to the case of a finite viscosity while still assuming that $\beta=0$. Viscosity tends to regularize the discontinuity in the solution in such a way that the SW front is smeared out on a finite time scale \be \delta \tau = \varkappa/ 2 \tilde q_0(\tau_0) \sim \varkappa/ \sqrt{F-F_{\rm cr}}.\ee The corresponding motion in the effective potential is illustrated in the Fig.\,\ref{Fig4}a. During the time interval $\delta \tau$ a particle propagates from the unstable point to a stable one under the action of the friction force specified in the Eq.~\eqref{q}. For sufficiently small viscosity, $\delta \tau \ll 1$, one can let $F \cos \tau \approx F \cos \tau_0$ within the front width. In this limit the Eq.~\eqref{q} is solved exactly with the result
\be
q(\tau)=\sqrt{2(F\!-\!F_{\rm cr})}\tanh\lt[{\sqrt{2(F\!-\!F_{\rm cr})}(\tau\!-\!\tau_0)}/{\varkappa}\rt],
\ee
which demonstrates that the smeared step is well described by the SW solution.

A simple analysis in the limit  $\varkappa \to 0$ yields the electric current $J$ and the dissipated power \be P=e \langle N V  E_0 \sin\theta \rangle\ee per unit volume [see Fig.\,\ref{Fig2}(e,f)]. In particular, we find
\begin{align}
&F<F_{\rm cr}:\,\,\, J\!=\!\frac{\pi^2F^2}{144 \omega F_{\rm cr}},\ \  P\!\equiv 0,\\
&F>F_{\rm cr}: \,\,\, J\!=\!\frac{4}{9\omega}\!\left( F\!-\!\frac{8 F_{\rm cr}}{\pi^2}\right)\!,\ \ P\!=\!C(F-F_{\rm cr})^{\frac{3}{2}},
\label{JP}
\end{align}
where $C=16\sqrt{2}  m N_0/81\pi$ is independent of viscosity. Remarkably, the current remains finite even for $\varkappa=\gamma=0$,  which implies that it has a diamagnetic nature. Even more interesting, the power $P$ remains finite above the threshold, $F>F_{\rm cr}$. In this regime, the energy dissipation occurs at the front of the SW in the region where the SW width is of the order of  $\varkappa$ and  is proportional to $\varkappa v \p^2v/\p \theta ^2 \propto 1/\varkappa$.  As the result  the viscosity $\varkappa$ drops out from the expression for the total dissipation \cite{Landau}. It is worth stressing that the strong-coupling result of Eq.~\eqref{JP} is essentially non-perturbative.

When SW does emerge, the behavior of $v_n$ qualitatively changes. This can be seen directly from the Fourier transform \be v=\sum_n v_n\exp(i n\theta).\ee For $F<F_{\rm cr}$, the high order harmonics decay exponentially with $n$ as $v_n \propto \exp(- a n)$, where $a \propto F_{\rm cr}-F$ at $F \to F_{\rm cr}$ (this estimate holds with an exponential precision).  Exactly at the threshold, one finds $v_n \propto 1/n^2$, while for $F>F_{\rm cr }$, the decay of harmonics $v_n$ is very slow, $v_n \propto 1/n$, which is the consequence of the step-like behavior of the solution [see Figs.\,\ref{Fig2}(c,d)]. This power-law dependence is valid for $n < 1/\delta\tau$.  Higher harmonics are exponentially suppressed due to the finite front width of the SW.  Hence, the generation of SW leads to a large increase of the excited harmonics and, consequently, to power dissipation.  

With increasing viscosity $\varkappa$  the SW front smears out and the non-linear oscillations are fully suppressed. The latter evolve into linear ones for $\varkappa \gg \sqrt F$.  The exact solution to Eq.~\eqref{q} for arbitrary $\varkappa$  (but $\beta=\gamma=0$) is presented in Appendix~\ref{appC2}.

\subsection{Generation of solitons due to the dispersion of plasma velocity}   

Let us now assume that $F>F_{\rm cr}$. We fix $\varkappa$ at sufficiently small value (such that $\delta \tau \ll 1$) and study what happens with increasing the dispersion coefficient  $\beta$. The solution is illustrated in Figs.\,\ref{Fig3}(a-d). We see that dispersion leads to a formation of solitons on the SW front. This process can again be understood by analyzing the mechanical analogy described above.

Since $\beta$ is responsible for ``inertia'' term in Eq.~\eqref{q} it is responsible for the transformation of a decaying solution (see Fig.\,\ref{Fig4}a) into an oscillatory one (see Fig.\,\ref{Fig4}b). For a finite, but sufficiently small $\beta$ ($\beta \ll \varkappa$), the SW front remains sharp so that one can still assume $F \cos \tau \approx F \cos \tau_0 $ within the front width. Then, the characteristic scales of the problem can be understood from the analysis of Eq.~\eqref{q} linearized near the stable point, \be \beta  \ddot{\delta q} +\varkappa  \dot{\delta q} +(\varkappa /\delta \tau)  \delta q=0,\ee where $\delta q=q-\tilde q_0(\tau_0)$. By looking for the solution in the form  $\delta q \propto \exp[\lambda \tau]$  we find  \be \lambda=-(\varkappa/2\beta)(1\pm\sqrt{1-\beta/\beta_{0}}),\ee where $\beta_0=\varkappa \delta \tau/4\ll \varkappa$.

For $\beta\ll\beta_0$, we find two exponentially decaying solutions. The slowest decay corresponds to $\lambda  \approx -1/\delta \tau$. In this solution the dispersion does not play an essential role as can be seen from Fig.\,\ref{Fig3}a (for simplicity, in Fig.\,\ref{Fig3} we consider $F\gg F_{\rm cr}$). We note that finite viscosity broadens the SW front on the scale of $\delta\tau$.

One can also see that the solitons start to build up for $\beta>\beta_0$. Indeed, in this case, the exponent $\lambda$ acquires an imaginary part hence the oscillations appear on top of the smeared wave front as shown in Fig.\,\ref{Fig3}b. (Note that similar effects also arise in the Luttinger liquids due to the same reason \cite{prot1,prot2}). For the case $\beta_0 \ll \beta \ll \varkappa$, we find two rapidly oscillating and slowly decaying solutions. The number of oscillations during the decay from unstable to stable point (the number of solitons $N$) can be estimated as the ratio of imaginary part of $\lambda$ to its real part that yields \be N \sim \sqrt{\beta/\beta_0}.\ee
The number $N$ increases with increasing $\beta$ until $\beta \sim \varkappa$ (see Fig.\,\ref{Fig3}c).

When $\beta$ becomes larger than $\varkappa$ both solutions do not decay for the whole oscillation period of the external force, $0<\tau<2\pi$.  As the result, the  viscosity can be fully neglected in the limit $\beta \gg \varkappa$. In this case the transition from unstable point to stable one is governed by an adiabatically slow variation of the potential. The number of oscillations in this limit can be estimated as  \be N\sim {\rm Im}\lambda \sim \varkappa/\sqrt{\beta \beta_0}.\ee The result of this equation is illustrated in Fig.\,\ref{Fig3}d. This analysis suggests that the maximal value of solitons is achieved for 
$\varkappa \sim \beta$ with \be N_{\rm max} \sim \sqrt{\varkappa/\beta_0} \propto 1/\sqrt{\delta \tau}.\ee

Different regimes are summarized in Fig.\,\ref{Fig3}e in the coordinates $(\varkappa,\beta)$. The dependence of $N$ on $\beta$ is plotted in Fig.\,\ref{Fig3}f.
A more detailed analytical study in the limit $\beta \gg \varkappa$ is presented in Appendix~\ref{appD}.

\section{ Discussion } \label{discussion}
\begin{figure}[t!]
\center{\includegraphics[width=0.9\columnwidth]{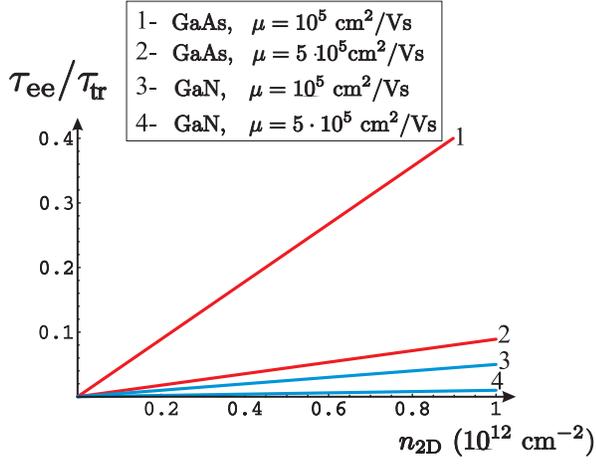}}
\caption{Dependence of parameter $\tau_{\rm ee}/\tau_{\rm tr}$  on the electron concentration for GaAs and GaN structures with different mobilities.}
\label{Fig5}
\end{figure}
\begin{figure}[t!]
\center{\includegraphics[width=0.9\columnwidth]{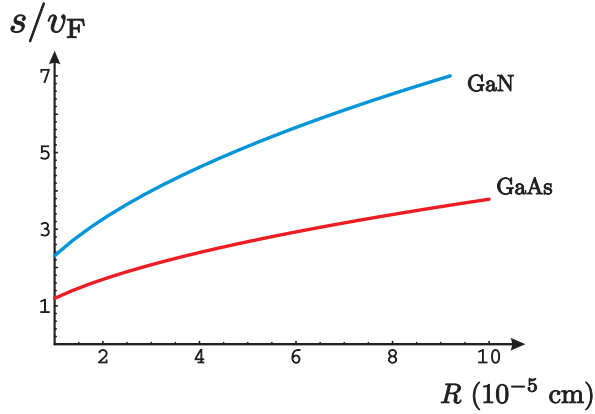}}
\caption{Ratio of the plasma wave velocity to the Fermi velocity for a 2D gas as a function of the nanoring radius.}
\label{Fig6}
\end{figure}
\begin{figure}[t!]
\center{\includegraphics[width=0.9\columnwidth]{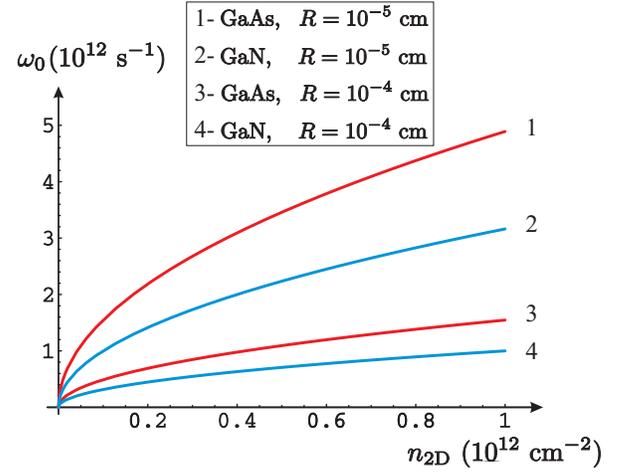}}
\caption{Dependence of the fundamental plasma frequency on the electron concentration for nanorings of different sizes.}
\label{Fig7}
\end{figure}
\begin{figure}[t!]
\center{\includegraphics[width=0.9\columnwidth]{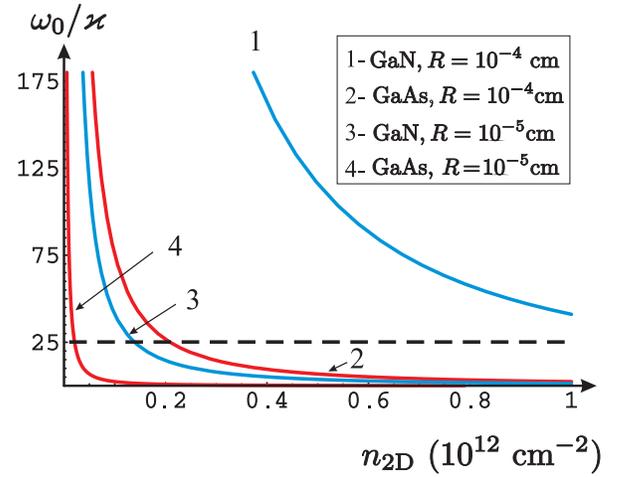}}
\caption{Viscosity-related quality factor plasmonic resonance as a function of electron concentration for the nanorings of different sizes. Dashed line corresponds to the case $\varkappa=\beta$,  where number of solitons in nonlinear regime is maximized. Above this line all nonlinear solutions corresponds to multiple solitons (see also Fig.\,\ref{Fig3}e).                 }
\label{Fig8}
\end{figure}
\begin{figure}[t!]
\center{\includegraphics[width=0.9\columnwidth]{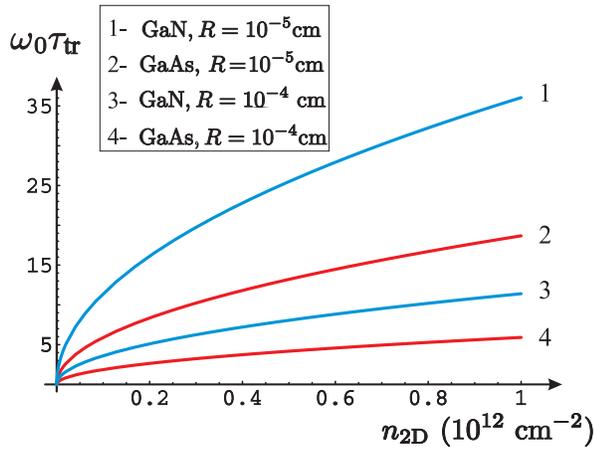}}
\caption{Quality factor of plasmonic resonance related to phonon and impurity scattering as a function of the electron concentration for nanorings of different sizes made of material with the mobility $\mu=10^5~$cm$^2$/Vs. }
\label{Fig9}
\end{figure}
\begin{figure}[t!]
\center{\includegraphics[width=0.9\columnwidth]{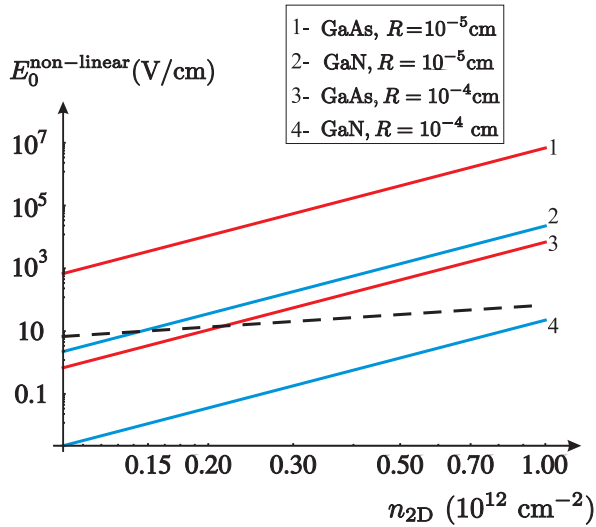}}
\caption{Characteristic value of radiation-induced electric field that is determined by the condition $\sqrt F =\varkappa$ (solid lines)  and  by the condition $\sqrt F =\beta$ (dashed line) as a function of the electron concentration. For a given $R$ the nonlinear regime corresponds to values of field that is above both the solid and the dashed line.}
\label{Fig10}
\end{figure}

Let us discuss the application of the model developed to realistic nanorings. The plasmonic resonances predicted above can be observed in 3D and 2D semiconductor and metallic rings as well as in ring arrays (see Fig.\,\ref{Fig1}). In particular, 2D rings, which are depicted in Fig.\,\ref{Fig1}b, can be fabricated by growing the standard 2D semiconductor or graphene layers followed by patterning gated or ungated nanorings or nanoring arrays. The estimates below show that the conditions needed for observing both linear and nonlinear plasmonic resonances can be easily met for a typical semiconductor at realistic temperatures. Let us present the detailed estimates for 2D GaAs and GaN nanorings. The main difference between these materials is due to the different effective masses: $m=0.067$  (in the units of  electron mass)  for GaAs and $m=0.2 $ for GaN.  

To be more specific let us choose the geometry relations
\be a=R/10,\qquad d= R/3. \label{geometry} \ee
for the rings with $10^{-5}$\,cm $<R < 10^{-4}$\,cm and $10^{11}$\,cm$^{-2} <n_{\rm 2D}<10^{12}$cm$^{-2}$ assuming that $T= 30^{\circ}$K.  For simplicity, we restrict ourselves to the ungated case such that the ring is placed on the surface between the air and a substrate with a dielectric constant $\varepsilon_1$  which is close to the dielectric constant of 2D layer. In this case $\varepsilon= (\varepsilon_1+1)/2$. Below we shell use $\varepsilon = 6.3$ for GaAs rings and $\varepsilon = 5$  for GaN rings.   

The parameter that ensures the validity of the hydrodynamic approach is the ratio of the electron-electron collision time to the momentum relaxation time, $\tau_{\rm ee}/\tau_{\rm tr}.$ In the hydrodynamic regime (electron collision-dominated) this parameter is small. The rough estimate of the collision time is given by $\hbar/\tau_{\rm ee} \sim T^2/E_{\rm F}$ \cite{comment1}. Expressing momentum relaxation time via the electron  mobility and Fermi energy via electron concentration, we find $\tau_{\rm ee}/\tau_{\rm tr} \sim \pi e\hbar^3 n_{\rm 2D}/(m^2T^2 \mu)$.  In Fig.\,\ref{Fig5}, we plot this parameter as a function of  $n_{\rm 2D}$ for two values of mobility: $\mu=10^5$\,cm$^2/$V s  and  $\mu=5 \cdot 10^5$\,cm$^2/$V s.  Larger value of the mobility is still well below a record mobility for 2D Ga As structures at such temperatures.   We see that the condition $\tau_{\rm ee}/\tau_{\rm tr} <1$ is satisfied even for the case of sufficiently low mobility value and the condition $\tau_{\rm ee}/\tau_{\rm tr} <1$ is satisfied for both materials in the whole range of available electron concentration.

The main advantage of the proposed system is a high operation speed that is defined by a particularly large value of the typical plasma wave velocity as compared to electron velocity. For rings prepared on the basis of 2D materials, the ratio $s/v_{\rm F}$ does not depend on electron concentration, $s/v_{\rm F}= \sqrt{e^2 a\Lambda m/(\pi\varepsilon \hbar^2)} $ [here we used Eq.~\eqref{s}].  Having in mind Eq.~\eqref{geometry}, one readily finds the dependence of the ratio $s/v_{\rm F}$ on the ring size (see Fig.\,\ref{Fig6}). It can be seen that for chosen parameters, the condition $s/v_{\rm F} >1$ is satisfied.

The  dependence of the fundamental  plasma frequency on the electron concentration is shown in Fig.\,\ref{Fig7}.    In the chosen interval of the electron concentrations and the ring sizes, the plasmonic frequency is in the terahertz range. Hence,  the  proposed  ring-based  devices  are very attractive for possible applications in terahertz electronics and optics.

To discuss possible experimental realizations let us estimate nanoring quality factors. It follows from the Eq.~\eqref{rescaled2} that in a vicinity of the resonance, $\omega-\omega_0 \ll \omega_0$, the ratio $\omega_0/\beta$ is determined by a geometrical factor,
\be
\frac{\omega_0}{\beta} \simeq  \frac{R^2 \Lambda}{d^2} \approx 25,
\label{omega/beta}
\ee
where the result of Eq.~\eqref{geometry} is taken into account. The viscosity in this regime is estimated as
\be
\varkappa =\frac{\eta}{R^2 } \sim \frac{v_{\rm F}^2 \tau_{\rm ee}}{R^2} \sim  \frac{2\pi^2\hbar^5 n_{\rm 2D}^2}{R^2  m^3 T^2},
\label{varkappa}
\ee
hence the viscosity-related quality factor turns out to be large to the extent that the viscosity does not suppress plasma resonances. Indeed, the solid lines in the Fig.\,\ref{Fig8} show the viscosity-related quality factor $\omega_0/\varkappa$  as a function of the electron concentration for nanorings of different sizes. The dashed line corresponds to the case $\varkappa=\beta$, where the number of solitons $N$ is maximal.  Above this line all nonlinear solutions would correspond to the regime of multiple solitons (see also Fig.\,\ref{Fig3}e). The shock wave solutions take place in the region that is well below this line. 

In order to demonstrate that momentum relaxation due to disorder and phonons does not destroy the plasmonic resonances, we plot in Fig.\,\ref{Fig9} the corresponding quality factor $\omega_0\tau_{\rm tr}$ as a function of the electron concentration. In this plot we substitute a relatively low value of electron mobility, $\mu=10^5$\,cm$^2$/Vs. Still, even for such a value, the disorder-related quality factor remains sufficiently large for typical electron concentrations. Since the quality factor is simply proportional to the mobility, the use of samples with higher mobility, e.\,g. $\mu=5\cdot 10^5$\,cm$^2$/Vs (which is still well below the record mobility value for 2D GaAs) would lead to the fivefold  enhancement in the quality factor as compared to the numbers presented in Fig.\,\ref{Fig9}. Thus, for realistic parameters of a semiconductor nanoring, the combined quality factor of the plasmonic resonance is certainly large enough to make the proposed physics plausible. 

Before closing the Section let us briefly discuss the conditions that need to be met in order to observe the non-linear regime. Exactly at the resonance ($\delta=0$) the non-linear behavior occurs for $\sqrt{F} >\{\varkappa, \beta\}$ (see Fig.\,\ref{Fig3}e). These conditions can be, respectively,  rewritten as
\begin{eqnarray}
e E_0  &>&   2mR\varepsilon \varkappa^2/3,\\
\label{cond1}
e E_0 &>&  2m R\varepsilon  \beta^2 /3 \simeq 2m R\varepsilon(\omega_0 /25)^2/3,
\label{cond2}
\end{eqnarray} 
where we took advantage of the result of the Eq.~\eqref{omega/beta}. With the help of the Eqs.~(\ref{omega0},\ref{varkappa}) we obtain a characteristic field that is required to observe the nonlinear regime, $E_0^{\rm non-linear}$. This field must larger than the fields at the right hand sides of the inequalities \eqref{cond1} and \eqref{cond2}. The non-linear plasmonic excitations, i.\,e. the solitons and the SWs, is, therefore, expected to form for $E_0>E_0^{\rm non-linear}$.

The r.\,h.\,s. of the inequality \eqref{cond1} is plotted in Fig.\,\ref{Fig10} with the solid lines as a function of the electron concentration in a nanoring. Similarly, the r.\,h.\,s. of  the Eq.~\eqref{cond2} is plotted in the same figure with the dashed line (one can check, indeed, that in view of the Eq.~\eqref{geometry}, the r.\,h.\,s. of the Eq.~ \eqref{cond2}  does not depend on the ring radius $R$). Thus, for a given $R$, the nonlinear regime takes place for the fields $E_0$ that stay above both the solid and the dashed line. 

Let us consider, for example, a GaAs ring with the radius $R=10^{-4}$\,cm (the curve $3$ in the Fig.\,\ref{Fig10}).  Intersection of this curve with the dashed line occurs at the concentration $n_{\rm 2D}^*\approx 0.25 \cdot 10^{12}$\,cm$^{-2}$. The non-linear regime is, therefore, realized for the values of $E_0$ that are above the dashed line, provided $n<n_{\rm 2D}^*$. In this regime $\beta>\varkappa$ hence our theory predicts multiple soliton solutions. We see that the corresponding value of $E_0^{\rm non-linear}$ is sufficiently small and can be achieved in experiment. 

For $n>n_{\rm 2D}^*$  the non-linear regime is realized for the values of $E_0$ that are larger than the values given by the curve 3 in the Fig.\,\ref{Fig10}. In this case, $\beta<\varkappa$ hence our theory predicts the SW solutions as well as solitons developed at the front of the shock wave (see Fig.\,\ref{Fig3}e). The value of $E_0^{\rm non-linear}$ in this case is larger or about the value $10^2 \div10^3$V/cm (depending on the electron concentration). Such a value can be easily reached in modern sources of GHz and THz radiation.

We should also mention that the non-linear regimes discussed above is even easier to realize with the pulsed source of radiation (the minimum pulse width is only limited by the period of the electromagnetic wave). Since the \textit{dc} current and the induced magnetic moment arise due to rectification of alternating electric field the entire analysis applies to this regime of operation as well \cite{comment}. Thus, even for nanorings made of GaAs of smaller radii (see the curve 1 in Fig.\,\ref{Fig10}), the nonlinear regime can be realized provided that the electron concentration is not too large. 

Finally, we should estimate the magnetic field induced by the current circulating in the ring. For GaAs ring with $n_{2D} =10^{12}$\,cm$^{-2}$ and $R =10^{-5}$\,cm subject to radiation with $E_0=10^4$V/cm we find the circulating  \textit{dc} current that is given by $1.5$\,$\mu$A and the magnetic field in the center of the ring that is given by $0.1$\, Gauss.

\section{Conclusion} \label{conclusion}

To conclude, we demonstrate that a circularly-polarized radiation may induce a strong diamagnetic \textit{dc} current in a nanoring, which is dramatically enhanced in the vicinity of plasmonic resonances. When the amplitude of external field exceeds a critical value $F_{\rm cr}$, shock waves and/or solitons are formed. In this regime the current and magnetic moment grow linearly with the amplitude of the external field and a large number of the THz-frequency harmonics can be generated by the device. We demonstrate that the effect can be observed in nanorings made of 2D semiconductors in the standard range of electron concentrations and for realistic ring sizes. The quality factor of the resonances can be as high as $10 \div100$. The amplitude of the exciting wave driving the system into the non-linear regime is shown to be not too large, of the
order of $10\div10^3~$V/cm. The effects can be easily scaled up by preparing the arrays of nearly identical nanorings. 

The discovered enhancement of the diamagnetic current by plasmonic resonances should enable numerous applications of ballistic nanorings and nanoring arrays including, but not limited to the electric field control of magnetic forces and the new ways to construct highly efficient low-loss switches that operate in a wide range of frequencies from microwave to the upper bound of the THz range.

\begin{acknowledgments}

We thank I.\,Gornyi, A.\,Kimel,  A.\,Mirlin, D.\,Polyakov, and I.\,Protopopov for stimulating discussions. The work of M.T. was supported by the EU Network FP7-PEOPLE-2013-IRSES Grant No 612624 ``InterNoM'' and by Dutch Science Foundation NWO/FOM 13PR3118. The work of K.L.K. and V.Yu.K. was supported by Russian Science Foundation (grant No. 16-42-01035)

\end{acknowledgments}

\appendix

\section{Electrostatic potential}
\label{appA}
We start by deriving the Eq.~(\ref{Phi}) of the main text. Let us consider electrostatic force (per unit mass) created by electrons distributed along the ring with the concentration $N=N(x)$. We assume that the Coulomb interaction is screened on the scale $d$ such that $d \ll R$. Then, in the limit of infinitely thin wire, the force per unit mass acting on the electric flow at the point $x$ can be written as $-{\p \Phi}/{\p x}$, where
\be
\Phi= \frac{e^2 }{m\varepsilon} \int dx'  [N(x')-N_0] \frac{\exp(-|x-x'|/d)}{|x-x'|}.
\label{force}
\ee
The integral entering Eq.~\eqref{force}  diverge logarithmically  at $x' \to x$.  This divergency is regularized by taking into account a finite width $a$ ($a\ll d$) of the ring.  Assuming that $N(x)$ changes slowly on the scale $d$ we may cast the electron concentration in the form $N(x') \approx N(x)+ N'(x) (x-x')+(1/2)N''(x) (x-x')^2$. Substituting this equation into Eq.~\eqref{force} and performing (with logarithmic precision) the integration over $x'$ we arrive at Eq.~\eqref{Phi} of the main text.

\section{Linear solution for finite friction at the surface}
\label{appB}
In this section, we briefly discuss the effect of the surface friction in the linear regime, i.\,e. for the linear plasmonic excitations.

The surface friction leads to a inhomogeneous distribution of the velocity and concentration in the radial direction. In the resonance approximation, linearized velocity can be written as $v=v_1(r)\exp(i \theta) +{\rm h.\,c.}$, where $v_1$ yields the equation
\be
v_1[i(2\delta-\beta)+\varkappa +\gamma] -\varkappa  \frac{R^2}{r}\frac{\p}{\p r}\left( r \frac{\p v_1}{\p r}\right)=\frac{eE_0}{2 i m R\varepsilon}.
\label{v1}
\ee
Here $r$ is the radial coordinate such that $0<r<a$.  Since our calculations have illustrative character, we do not distinguish here between bulk and shear viscosity, characterizing the electron liquid by a single viscosity coefficient $\varkappa.$  We further assume that the friction force is proportional to the velocity and can be modeled by a delta-function potential  on the surface of the ring, $V f \delta(r-a)$, where $f$ is a certain coefficient. In this model we find the boundary condition to Eq.~\eqref{v1} as
\be
\left.\frac{\p v_1}{\p r}\right|_{r=a}= - k v_1,
\label{bound}
\ee
where $k=f/\eta.$  For sufficiently large radius $R$ such that $a^2\lt|i(2\delta-\beta)+\varkappa +\gamma\rt|/\varkappa R^2\ll1$,
the solution to Eq.~\eqref{v1} with the boundary condition of Eq.~\eqref{bound} reads
\be
v_1\approx\frac{eE_0}{2 i m R\varepsilon} ~\frac{1 +k (a^2-r^2)/2a}{\tilde \varkappa +\gamma +i(2\delta -\beta)},
\ee
where $\varkappa =\varkappa (1+ 2k R^2/a)$. For the limit
\be
k \ll \frac{a}{R^2},
\label{k}
\ee
or, equivalently, for $f\ll \varkappa a$, we restore Eq.~\eqref{vv} of the main text.  Hence, the inequality \eqref{k} represents a criterium for neglecting the friction force. For lager values of $k$ the friction would modify our results. As far as $ a/R^2 \ll k \ll 1/a$ the modification is simply reduced to replacing $\varkappa$ in  Eq.~\eqref{vv} with a large constant $\tilde \varkappa \gg \varkappa$.  For even larger values, $k \gg 1/a$, one obtains the dynamical Poiseuille flow in which velocity goes to zero on the nanoring surface.

\section{Numerical solution of hydrodynamic equations}
\label{appE}
In this section, we analyze the most general case of non-stationary hydrodynamic equations in the presence of dispersion, viscosity, and disorder-induced friction. In the rotating reference frame ($t'=t$ and $\theta=\varphi-\omega t$), these equations read
\begin{align}
\label{nonst1new}
&\frac{\p n}{\p t'}+\frac{\p}{\p\theta } \left[(1+n) v - \omega n \right]=0,\\
&\frac{\p v}{\p t'}\!+\!\frac{\p}{\p\theta}\left[   \frac{v^2}{2}\!-\!v\omega\!+\!\omega_0^2 n\!-\!\varkappa \frac{\p v}{\p \theta}\!+\!\beta \frac{\p^2 v}{\p \theta^2} \right]= \!-\gamma v\!+\!\frac{eE_0}{mR}\sin\theta,
\label{nonst2new}
\end{align}
For the resonance approximation, $\delta \ll\omega_0$, the solution to these equation is very close to a stationary solution in the rotating reference frame. In the other words, we may assume that derivatives $\p/\p t'$ are on the order of $\delta$ and, therefore, are small compared to $\omega$. Then, Eqs.~\eqref{nonst1new} and \eqref{nonst2new} can be somewhat simplified. As the first step we rewrite Eq.~\eqref{nonst1new} as
\be
\frac{\p n}{\p \theta}=\frac{1}{\omega} \left[   \frac{\p n}{\p t' }+\frac{\p (1+n)v}{\p\theta}\right].
\ee
In the next step we substitute $n\approx  v/\omega$ into the r.\,h.\,s. of this equation.  As a result, we obtain a closed non-stationary equation for the velocity
\be
2\frac{\p v}{\p t'}\!+\!\frac{\p}{\p\theta}\left[\frac{3}{2}v^2\!+\!2\delta v\!-\!\varkappa \frac{\p v}{\p \theta}\!+\!\beta \frac{\p^2 v}{\p \theta^2} \right]=\!-\gamma v\!+\!\frac{eE_0}{mR}\sin\theta,
\label{nonst}
\ee
which is easily solved by the standard built-in realization of the finite-element method in Mathematica. For sufficiently small $\gamma$ and for $\varkappa \ll \sqrt F$, $\beta \ll \sqrt F$ we find the solution to be stationary in the rotating reference frame at sufficiently long times. This reproduces the results that are shown in Figs.\,\ref{Fig2} and \ref{Fig3}. Also, in the limit $\beta_0<\beta <\varkappa$ the numerical simulations yield the values of $\alpha$ and $\varepsilon_0$, which are in a very good agreement with those  found analytically  [see Eq.~\eqref{aeps} below].

\section{Exact solutions}
\subsection{Exact solution at $F\ll F_{\rm cr} $ and  $F>F_{\rm cr}$  for $\gamma=\varkappa=\beta=0$.}
\label{appC1}
The linear regime is analyzed by expanding Eqs.~\eqref{q-tau} and \eqref{aver} in $F$. Simple analysis yields
\be
q_0\approx \delta+\frac{F^2}{16\delta^3},\qquad q\approx  \delta-  \frac{F\cos\tau}{2\delta}.
\ee
Substituting $v=2(q-\delta)/3$ we get
\be
v\approx -\frac{F\cos\theta}{3\delta},
\label{vv1}
\ee
that should be compared to Eq.~\eqref{vv} of the main text for $\gamma=\varkappa=\beta=0$.

With increasing value of $F$ the absolute value of $q_0$ also increases.  When $F$ reaches the critical point $F=F_{\rm cr}$ the value of $q_0$ is given by $|q_0|= \sqrt F=\sqrt{F_{\rm cr}}$.  At this point the positive and the negative solution of Eq.~\eqref{q-tau} read
\be
q(\tau)= \pm \sqrt{2F} \left|\sin\left(\frac{\tau}{2}\right)\right|= \pm\frac{\pi }{ 2} \left|\delta \sin\left(\frac{\tau}{2}\right)\right|,
\label{qq-tau}
\ee
while the velocity is given by
\be
v= \frac{2\delta}{3} \left( \frac{\pi}{2}\left|\sin\frac{\theta}{2} \right|-1 \right),\qquad \mbox{for}  \quad F= F_{\rm cr}
\label{vv2}
\ee
It is evident from Eq.~\eqref{qq-tau} that at $F= F_{\rm cr}$ the positive and the negative solution touch each other at the points $\tau=0$  and $\tau=2\pi$. At these points one finds $\tilde q_0=0$ and the positions of extrema coincide, hence there appear a possibility to jump between the two solutions from the stable point to the unstable one. With $F$ increasing above $F_{\rm cr}$ the Eq.~\eqref{aver} of the main text no longer has any continuous solution. Therefore, for $F>F_{\rm cr}$, one should search for a solution that is discontinuous: $q = -\tilde q_0 (\tau) $ for  $0<\tau<\tau_0$ and  $q = \tilde q_0 (\tau) $ for $\tau_0<\tau<2\pi$. At the discontinuity point $\tau=\tau_0$ there exists a jump from the positive solution to the negative one. The negative solution changes back to the positive one at $\tau=2\pi$ so that the periodicity condition is fulfilled.

The discontinuity position is fixed by the condition $\langle q\rangle=\delta$ that is written as
\be
-\int_0^{\tau_0}d\tau \tilde q_0(\tau)+\int^{2\pi}_{\tau_0}d\tau \tilde q_0(\tau)=\delta.
\ee
Integrating the latter one finds Eq.~\eqref{cos} of the main text.

For $F> F_{\rm cr}$ the velocity reads
\be
v= \frac{2}{3}\BC \sqrt{2F} \sin(\theta/2) -\delta,& 0<\theta<\theta_0, \\
-\sqrt{2F} \sin(\theta/2) -\delta,& \theta_0<\theta<2\pi,
\EC
\label{vv3}
\ee
where the angle $\theta_0=-\tau_0$ obeys the relation $\cos(\theta_0/2)= \sqrt{F_{\rm cr}/F}$.

\subsection{Exact solution for $\gamma=\beta=0$ and arbitrary $\varkappa$.}
\label{appC2}
In the absence of dispersion ($\beta=0$) the Eq.~\eqref{q} simplifies to
\be
\varkappa \dot{q}=q_0^2-q^2-F\cos\tau.
\label{q1}
\ee
With the help of new variables
\be
z=\tau/2,\qquad q=\frac{\varkappa}{2 y} \frac{dy}{dz},
\label{vary}
\ee
we rewrite Eq.~(\ref{q1}) in the canonical form of the Mathieu equation
\be
\frac{\p^2 y}{\p \varphi^2}+[a-2Q\cos{(2\varphi)}]y=0,
\label{Mathieu}
\ee
where  $a=-4\langle q^2\rangle/\varkappa^2$ and $Q=-2F/\varkappa^2$.

The constraint $\langle q\rangle=\delta$ can be rewritten in terms of the function $y(z)$ as $y(\pi)=\exp{(2\pi\delta/\varkappa)}y(0)$.
Thus, we get
\be
\mu(a,Q)=2i\delta/\varkappa,
\label{mu}
\ee
where $\mu(a,Q)$ is the Mathieu characteristic exponent. The parameter $a$ is not a free external parameter but, in fact, has to be found self-consistently by calculating the average $\langle q^2\rangle$.  Instead of direct calculation of the average one may simply use Eq.~\eqref{mu}, which implicitly defines the dependence $a(Q,\delta)$.

Following this route one can express $y$ in terms of the solution of Mathieu equation as follows
\begin{align}
y(z) = &G[a(Q,\delta) ,Q,z]= \nonumber\\
 &{\rm MCos}[a(Q,\delta),Q,z] - i{\rm  MSin}[a(Q,\delta),Q,z],
\label{y1}
\end{align}
where ${\rm MCos}$ and ${\rm MSin}$ are Mathieu cosine and sine, respectively.  Using Eq.~\eqref{y1} one can readily express the velocity $v$ in terms of the angle $\theta$ as
\be
v(\theta)=\!-\frac{2}{3}\delta\!+\!\frac{2\varkappa}{3}  \frac{\p \ln[G[a(-2F/\varkappa^2,\delta),\!-2F/\varkappa^2,\!-\theta/2]}{\p\theta}.
\ee
The numerical analysis of this equation allows one to reproduce various regimes shown in Figs.\,\ref{Fig2} and \ref{Fig3}.  In the limit $\varkappa \to 0$ we recover analytical solutions obtained above [see Eqs.~\eqref{vv1}, \eqref{vv2}, and  \eqref{vv3}].

\section{Description of solitons in the limit $\beta \gg\varkappa$}
\label{appD}
For $\beta \gg \varkappa$ the viscosity can be fully neglected. Let us consider the electron dynamics assuming for simplicity that $\delta=0$ and, as a consequence, $F_{\rm cr}=0$.  In this case $\delta \tau \simeq \varkappa /\sqrt F$ and $\beta_0 \simeq \varkappa^2/\sqrt F$.  We assume that $\varkappa \ll \sqrt F$ hence $\delta \tau \ll 1$.  If the potential $\tilde U(q)$ were static the electron energy would conserve. In fact, the potential slowly changes due to the variation of $\tilde q_0$,  so that electron undergoes fast oscillations with the frequency of the order of  $\varkappa /\sqrt{\beta \beta_0} \sim  \sqrt{\varkappa/\beta \delta \tau} \sim F^{1/4}/ \beta^{1/2}$,  while its energy changes adiabatically.

\begin{figure}[t]
\center{\includegraphics[width=0.9\columnwidth]{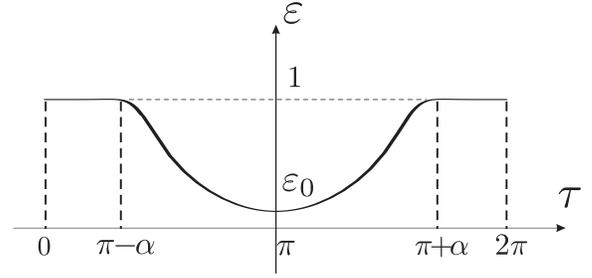}}
\caption{Dependence of $\varepsilon$ on $\tau$ for $ \varkappa<\beta$.  }
\label{Fig11}
\end{figure}

Let us discuss this process in more detail.  First, we consider what happens on the short time scales that are much shorter than the period of the external force.  We introduce a dimensionless coordinate $z$ and the energy $\varepsilon$: $E=(2\tilde q_0^2/3)\varepsilon$, $q=\tilde q_0 z$. Stable and unstable points of the potential correspond to $\varepsilon=-1$ and $\varepsilon=1$, respectively.   Frequency of the electron oscillations in the potential depends on $\varepsilon$:  $\omega =\sqrt{2\tilde q_0/\beta } ~\Omega (\varepsilon)$ where
\be
\frac{2\pi}{\Omega(\varepsilon)}= 2 \int\limits_{z_1}^{z_2} \frac{dz}  {H(\varepsilon,z)} \simeq
\BC
2\pi,& \varepsilon =-1, \\
-\ln(1-\varepsilon),& \varepsilon  \to 1.
\EC
\label{Omega}
\ee
Here $H(\varepsilon,z)=\sqrt{2\varepsilon/3 +z-z^3/3}$ and $z_{1,2}=z_{1,2}(\varepsilon)$ are the turning points of the potential. The averaged value of the electron coordinate reads
$
\langle q \rangle_\omega= \tilde q_0 f(\varepsilon),
$
where
\be
f(\varepsilon)= \frac{ \int\limits_{z_1}^{z_2} {dz  z}  H^{-1}(\varepsilon,z)}{\int\limits_{z_1}^{z_2} {dz}  {H^{-1}(\varepsilon,z)}}
\BC
1,&\varepsilon =-1, \\
-1,& \varepsilon  = 1,
\EC
\ee
and $\langle \cdots  \rangle_\omega$ stands for the averaging over the fast oscillations with the frequency $\omega(\varepsilon)$. Simple numerical analysis shows that $f(\varepsilon)$ is very well approximated by $f(\varepsilon) \approx -1+2^{6/7} (1-\varepsilon)^{1/7}$.

Next, we study slow dynamics caused by a time dependence of $\tilde q_0$. In this case it is useful to define an adiabatic invariant $J(\varepsilon) =\oint p dq = J(1)j(\varepsilon)$, where $J(1)=(24 \sqrt {2}/5) \sqrt \beta \tilde q_0^{5/2}$ and
\be
j(\varepsilon)=\frac{5}{12} \int\limits_{z_1}^{z_2} {dz}  {H(\varepsilon,z)}\simeq
\BC
\frac{5\pi(1+\varepsilon)}{36}, \qquad \varepsilon  \to -1,\\
1, \qquad \varepsilon  = 1.
\EC
\ee
Numerically one can approximate $j(\varepsilon)  \approx   (1+\varepsilon)/2$.

We parameterize $q_0 = A F$, where $A$ is a dimensionless constant hence $\tilde q_0 \approx\sqrt{F}\sqrt{A-\cos(\tau)}$.  We also parameterize the energy at the time $\tau=\pi$ as $\varepsilon_0$. From the conservation of the adiabatic invariant we, therefore, conclude that the dependence of energy on time is implicitly given by the following equation
\be
[A-\cos (\tau)]^{5/4}j(\varepsilon)=[A+1]^{5/4}j(\varepsilon_0).
\label{ad1}
\ee
The dependence of $\varepsilon$ on $\tau$ that follows from Eq.~\eqref{ad1} is shown in Fig.\,\ref{Fig11}.

\begin{figure}[t]
\center{\includegraphics[width=0.7\columnwidth]{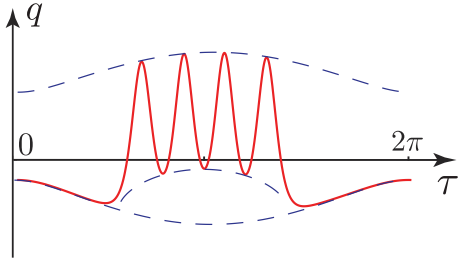}}
\caption{Numerical simulation of oscillations of $q$ for $\varkappa<\beta$ (red solid lines). Analytically calculated smooth envelopes are shown by dashed lines.}
\label{Fig12}
\end{figure}
At $\tau \approx \pi \pm\alpha$ the energy approaches the limiting value $\varepsilon =1$ and sticks to this point because $\Omega (1) =0$ [see Eq.~\eqref{Omega}]. In this regime the value of $q$ is given by $-\tilde q_0$.  From Eq.~\eqref{ad1}, we find the relation between $\alpha$ and $j(\varepsilon_0)$ as
\be
j(\varepsilon_0)\approx \left[\frac{A+\cos (\alpha)}{A+1}\right]^{5/4}.
\label{eps0}
\ee
At this point we have to take advantage of the constraint $\langle q\rangle =0$. To find the value of $\langle q\rangle$ one should average $\langle q\rangle_\omega $ over slow oscillations of the external field. This yields the following condition
\be
\int \limits_\pi^{\pi+\alpha}d\tau \tilde q_0(\tau) f[\varepsilon(\tau)]+   \int \limits_{\pi+\alpha}^{2\pi} d\tau [- \tilde q_0(\tau)]=0,
\label{ad2}
\ee
that allows one to determine $\alpha$. In particular, replacing the functions $f(\varepsilon)$ and $j(\varepsilon)$ in Eqs.~\eqref{ad1} and \eqref{ad2} with the corresponding approximative formulas, one arrives at the following equation for $\alpha$
\BEA
\int_0^{\alpha}\!dx \sqrt{A+\cos x}\!\left\{\!-1\!+\!2\!\left(\!1\!-\!\left[\frac{\!A\!+\!\cos(\alpha)}{A\!+\!1}\right]^{{5}/{4}} \right)^{{1}/{7}}  \right\}  =\nonumber \\
= \int_\alpha^\pi dx \sqrt{A+\cos x}.\quad
\label{ad3}
\EEA
Once the value of $\alpha$ is found one can use Eq.~\eqref{eps0} to determine $\varepsilon_0$. Parameter $A$ can be obtained from the numerical solution of hydrodynamical equations, $A\gtrsim 1$. Simple numerical analysis of Eq.~\eqref{ad2} yields
\be
\alpha \approx 1.3, \qquad \varepsilon_ 0 \approx 0.13.
\label{aeps}
\ee
The qualitative behavior of the function $q(\tau)$ is illustrated in Fig.\,\ref{Fig12}.

The values given in Eq.~\eqref{ad3} appear to be in a very good agreement with the solution obtained by direct numerical simulation of the original hydrodynamic equations. Below, we briefly describe the numerical method.


\begin{thebibliography}{99}
\bibitem{comment}
To find analytical expression for induced dc current in the regime of small pulse amplitudes, one should first expand the pulse into Fourier series, then find the linearized solution of the hydrodynamic equations, and finally substitute the obtained solution into Eq.~\eqref{averaged-I}, where the averaging should be now taken over the period of the pulse repetition.
\bibitem{comment1}  Actually,  collision rate contains also coupling constant $\hbar/\tau_{\rm ee} \sim  \alpha^2  T^2/E_{\rm F},$
where $\alpha \propto  e^2/(\hbar v_{\rm F} \varepsilon).  $   For typical parameters used for estimates, $\alpha \gtrsim 1.$ Here, for simplicity,
we assume $\alpha \sim 1 $ that yields low bound for collision rate.
\end{thebibliography}
\end{document}